# Hydrocarbon Contamination in Ångström-scale Channels

Ravalika Sajja,[a,b,†] Yi You,[a,b,†] Rongrong Qi,[a,b] Goutham Solleti,[a,b] Ankit Bhardwaj,[a,b] Alexander Rakowski,[c] Sarah Haigh,[c] Ashok Keerthi,[b,d] Boya Radha,[a,b,*]

[a]Department of Physics and Astronomy, School of Natural Sciences, The University of Manchester, Oxford Road, Manchester M13 9PL, United Kingdom

[b]National Graphene Institute, The University of Manchester, Oxford Road, Manchester M13 9PL, United Kingdom

[c]Department of Materials, School of Natural Sciences, The University of Manchester, Oxford Road, Manchester M13 9PL, United Kingdom

[d]Department of Chemistry, School of Natural Sciences, The University of Manchester, Oxford Road, Manchester M13 9PL, United Kingdom

[†] These authors contributed equally.

*correspondence to: radha.boya@manchester.ac.uk

## Abstract

Nonspecific molecular adsorption like airborne contamination occurs on most surfaces including those of 2D materials and alters their properties. While the surface contamination is studied using a plethora of techniques, the effect of contamination on a confined system such as nanochannels/pores leading to their clogging is still lacking. We report a systematic investigation of hydrocarbon adsorption in the angstrom (Å) slit channels of varied heights. Hexane is chosen to mimic the hydrocarbon contamination and the clogging of the Å-channels is evaluated *via* a Helium gas flow measurement. The level of the hexane adsorption, in other words, the degree of clogging depends on the size difference between the channels and hexane. A dynamic transition of the clogging and revival process is shown in sub-2 nm thin channels. Long-term storage and stability of our Å-channels is demonstrated here up to three years, alleviating the contamination and unclogging the channels using thermal treatment. This study highlights the importance of the nanochannels' stability and demonstrates self-cleansing nature of sub-2 nm thin channels enabling a robust platform for molecular transport and separation studies. We provide a method to assess the cleanliness of the nanoporous membranes, which is vital for the practical applications of nanofluidics in various fields such as molecular sensing, separation and power generation.

## Introduction

Nanopores and nanochannels play an important role both in fundamental studies of confined molecular transport[1], mimicking biological channels, as well as for applications in molecular and ion sieving membranes[2], sensors for biomolecular translocation[3], power generation[4], gas separation[5] and storage[6]. Some of the well-studied nanochannel systems include carbon nanotubes[7, 8], two-dimensional (2D) laminates such as graphene oxide[9], clays[10], MXenes[11], and quasi-zero dimensional pores such as atomic vacancies[12], nanopores punctured through 2D-materials[4, 13], to name a few. As much as the pore size is a critical factor in molecular transport for such systems[14], the influence of the nanochannel surface also becomes significant on mass transport when the pore is of sub-nanometre size, which is comparable to the length scales of molecular interactions[15].

Often, the surfaces, especially those of 2D-materials have high surface energy, and are inevitably adsorbed with unwanted molecules, resulting in the alteration of their properties[16]. For a long time, it was believed that graphite is hydrophobic as several measurements yielded water contact angles[17] around 90°. In 2013, Li *et al.*[18] reported the contact angles for freshly prepared graphene to be 44°, which increased to 80° after exposure to ambient atmosphere for a day. This change in graphene hydrophilicity was hypothesized to be caused by physisorption or chemisorption of the molecules (e.g. hydrocarbons) from ambient environment[19]. Such aging related contamination has also been found on other 2D materials such as $MoS_2$, $WS_2$ and InSe[20, 21]. The nature of the surface contamination on graphene was detailed by placing a layer of hexagonal boron nitride (hBN) on top, so that the contamination could be segregated into pockets due to the self-cleansing nature of 2D materials. These contamination pockets were examined by nanoindentation[22] as well as by cross-sectional transmission electron microscopy[23] to confirm that it is mainly composed of hydrocarbons. Even though the airborne hydrocarbon concentration is very low, ranging from parts per trillion to parts per million (ppm)[18], the alteration of the 2D materials' surface is prominent such as the variation of local carrier concentration[16] and surface change from hydrophilic to hydrophobic[18]. Several methods have been reported to drive off the contamination from 2D-material surfaces, such as dry cleaning using activated charcoal[24], thermal annealing[25] and polymer degradation by metal catalyst[26].

Basal planes of 2D-materials favour hydrocarbon adsorption and nanochannels/pores pose further challenging scenario as their edges are highly energetic and are decorated with functional groups, however only few reports address the hydrocarbon contamination in nanochannels[27]. Systematic study of the contaminants' impact on the nanochannels is essential, particularly to assess their stability and performance in the long run. Various techniques like ambient pressure x-ray photoelectron spectroscopy, tip-enhanced infrared spectroscopy, atomic force microscopy, and ellipsometry measurements, are used to characterise hydrocarbon contamination on surfaces[28-31]. However, those methods require sufficient concentration of hydrocarbons to be characterized whereas the contaminants in the nanochannels are far less and may be buried inside the channel limiting their access. Nevertheless, few attempts have been made to study the hydrocarbons under confinement, e.g., imaging of a passage of alkyl and alkenyl fullerenes inside a nanotube was performed using a transmission electron microscopy[32] and it was observed that imaging beam can itself cause changes in the hydrocarbons or activate free-radicals[32, 33]. Capillary condensation of hexane has been studied inside sub-10 nm pores, where phase changes leading to confinement induced vapor pressure elevation were observed[34]. Selective filling of hexane inside single wall carbon nanotubes was demonstrated[35], despite the diameter of the nanotubes being smaller than the kinetic diameter of hexane. Apart from the atmospheric contaminants, the flow of fluids in the confined channels can be altered by other contaminants such as polymeric residues arising from fabrication processes and by amorphous carbon deposits when using chemically synthesized 2D materials[36, 37].

Here we report gas flow measurements through channels made with angstrom-scale precision as a simple and easy way to track the hydrocarbon contamination under confinement. The channel surfaces are made from 2D materials serving as an ideal platform to understand hydrocarbon induced clogging or degradation pertaining to 2D material systems. We use helium gas flow measurement technique to check the airborne as well as deliberate hydrocarbon contamination using a model hydrocarbon, hexane.

## Results and discussion

In this work, we used slit-like channels of various heights ranging from ~ 0.4 nm to ~ 11 nm, fabricated from 2D materials namely graphene, and hBN. The fabrication recipe is described in our previous reports where we explored the ultrafast water flows and specular gas reflections off the surface of the walls[37, 38]. The slit-like channels are akin to sandwich of three 2D crystals named as bottom layer, spacer, and top layer, and these channels are assembled over a free-standing silicon nitride ($SiN_x$) membrane. Schematic of an Å-slit channel is illustrated in Fig. 1a and detailed fabrication procedure is presented in Fig. S1. Briefly, the bottom layer (~50 nm thick) and top layer (~200 nm) are mechanically exfoliated from the high-quality graphite or hBN crystals. Spacer layer is made from a thin layer of graphene which is cut out into parallel strips by electron beam lithography and plasma etching. The bottom and top layers define the channel walls. The distance between the spacer strips and thickness of the spacer layer determines the channel's width and height respectively. The channel height $h$ can be varied in multiples of numbers of layers $N$ of graphene spacer (shown in bottom inset of Fig. 1a). For example, single layer graphene spacer ($N$ = 1) means a ∼ 0.4 nm thick channel, and five-layer graphene spacer ($N$ = 5) corresponds to a ∼ 1.7 nm-thick channel. Cross-sectional image of a five-layer graphene slit is shown in Fig. 1b, where the individual layers of top and bottom atomically flat graphite crystals are clearly seen.

During and after the fabrication, the channels are susceptible to the polymeric and hydrocarbon contamination when stored in ambient conditions. Self-cleansing in sub-2 nm channels leads to such polymeric residues (usually, ∼ 1.5 to 2 nm) being squeezed out leaving cleaner channels, as examined in our previous study[37]. However, smaller hydrocarbon molecules can lead to channel clogging and degradation. We assess this *via* helium (He) gas flow measurement through the channels in vacuum. In the measurement setup (details in Fig. S2), He gas is input on one side of the device, whereas in the other chamber facing the exit of the channels, the output He is continuously quantified by a mass spectrometer. We monitor the gas flow rate for a given pressure difference and use it as an indicator of the channels' contamination level.

**Airborne contamination on Å-slit channels**

We first examine the clogging of a freshly fabricated device (Fig. 1c and Fig. 1d) with channels of height $h$ ∼ 1.7 nm (5-layer graphene spacer). The channels exhibit a noticeable reduction of helium permeance when the device is stored in ambient environment for 10 days, indicating a partial clogging. The degree of clogging is further exacerbated after an ambient air exposure of 20 days. Annealing is commonly used to decontaminate the graphitic carbon substrates as the fouled hydrocarbons are weakly bond through physisorption[39, 40]. It was found that upon annealing at high temperature (400 °C for 5 hours) in $H_2$/Ar atmosphere, the channels can be unclogged, and the (pristine state) flow can be recovered. Activated charcoal acts as a contamination sink and a good hydrocarbon adsorbent, as reported for dry cleaning of pristine graphene samples[24]. Hence, after annealing we stored the same device in charcoal for ∼50 days. A helium leak reduction of only ∼ 30 % suggests that indeed the observed clogging of channels is most likely due to the airborne hydrocarbon adsorption from the ambient environment.

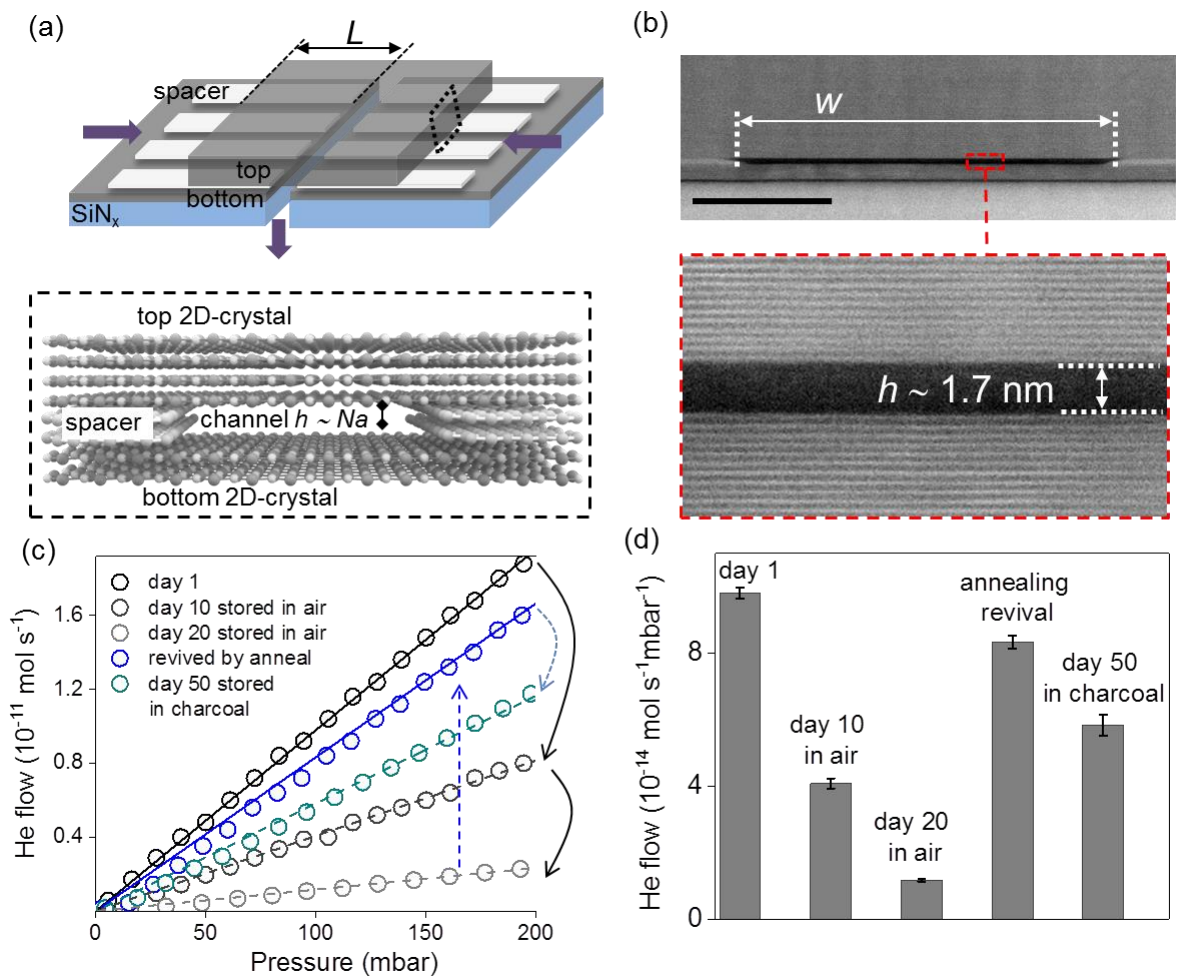

**Fig. 1** Airborne contamination in Å-channels. (a) Schematic of a device showing silicon nitride (SiN$_x$) membrane with angstrom slit channels on top, and their length $L$ is noted. Purple arrows indicate the flow directions of the gas through the device. The below inset shows a schematic representation of a channel displaying the top, bottom and spacer layers, with channel height $h$, labelled. $N$ is number of layers of graphene spacer, and $a$ is the interlayer distance in graphite. (b) Cross-sectional TEM dark field image of a 5-layer channel, with a magnified view shown below. Horizontal bright lines represent individual layers of graphite, and the dark space is the Å-channel. Scale bar of top image, 50 nm. (c) Helium flow through a Å-channels made with 5-layer graphene spacer (height, $h \sim 1.7$ nm), top and bottom graphite walls. The channels are ~ 5 μm long, and there were ~ 200 channels in the device. (d) Bar graph showing the data in (c) for the Å-channel device clogging and revival. Upon storage in ambient conditions, the flow reduced within few days, probably due to the channel clogging. However, the flow can be regained by annealing at high temperature. Storage in charcoal displayed a minimal flow reduction over ~ 50 days.

To verify if the hydrocarbon contamination commonly exists in other channels of different thicknesses, we performed the helium leak test as above, on devices with 3-layer and 32-layer channels, respectively. The gas permeability plots in Fig. 2 demonstrate that the extent of the clogging is not similar for these two devices. Specifically, the 3-layer device ($h \sim 1$ nm) has displayed $\sim 50\%$ of the He flow reduction, whereas the 32-layer device ($h \sim 11$ nm) showed more than $> 97\%$ reduction in the permeation and only a minor revival is seen upon the thermal treatment. Airborne contamination in a monolayer channel device monitored over few months is shown in supporting information Fig. S4. Similar to the 3-layer channel, the reduction in the flow over time through the channels could be revived by high temperature annealing (400 °C in $H_2$/Ar atmosphere). The difference observed in thick versus thin channel clogging could be related to interaction strength between hydrocarbon molecule and channel surface under varying confinement. Even a high temperature annealing (in combination with charcoal storage) is not efficient to revive the channels, implying a strong hydrocarbon- wall surface interaction inside thick channels. In absence of the self-cleansing effect, thick channels might be decorated with polymeric residues inside channels, which may seed further adsorption of the hydrocarbons.

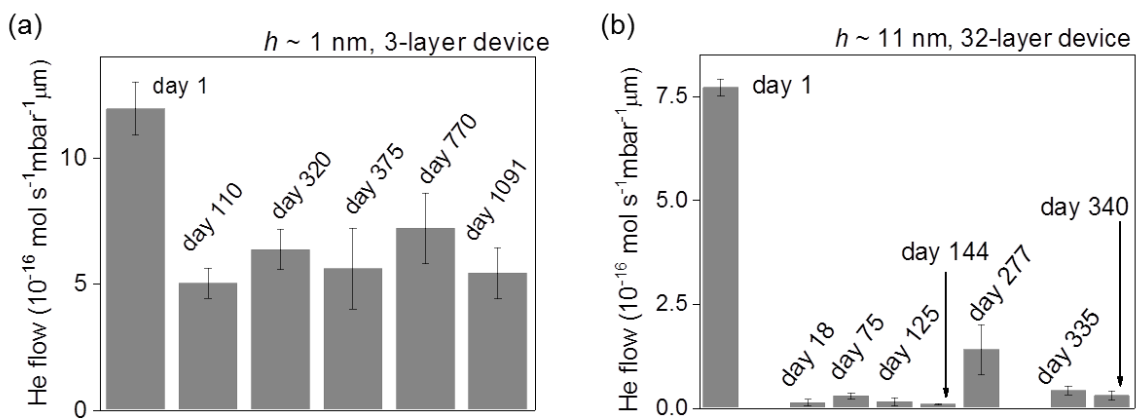

**Fig. 2** He flow through Å-channel devices with heights, (a) $h \sim 1$ nm (3-layer graphene spacer) and (b) h $\sim$ 11 nm (32-layer graphene spacer) monitored for over a duration of three years and one year, respectively. The bars represent He flows arbitrarily checked on different days through the same device when it was stored under charcoal. The devices were annealed before each measurement. Error bars are from two measurements on the same device.

Atmospheric hydrocarbons typically include combination of several alkanes, volatile organic compounds such as alcohols, aldehydes, ethers etc. In order to systematically probe the hydrocarbon molecule ingress into the channels, we chose hexane as a model molecule, due to three reasons, 1) it can be easily vaporized enabling experiments in the He flow measurement setup to monitor the effect of hexane exposure on the channels; 2) alkanes have a relatively strong adsorption and can form close packed layers on the graphene and hBN basal surfaces due to low potential energy and preferred orientation of carbons and hydrogens[41]; 3) its kinetic diameter is small enough (~ 4.2 Å), which allows us to conveniently choose Å-slit channels which are smaller and larger than its size to investigate the contamination. We have custom made our He flow measurement set-up integrated with a hexane injector as depicted in Fig. S2. By deliberately introducing the hexane, we examine the role of the height of the channels, in other words, height of the confinement on the hydrocarbon contamination and the ease of channel clogging and unclogging. Within few seconds of introduction of hexane into vacuum, saturated pressure of ~ 200 mbar is achieved, implying that the hexane is in a gas phase when approaching the nanochannels rather than the liquid form (Fig. S5). Control experiments with an aperture of diameter ~ 60 nm, shows that there is no unintended influence of hexane vapors on the measurement setup (further details in section S6). Vaporized hexane in vacuum is passed through the channels in vacuum, and the gas conductance of He through channels before and after Hexane exposure is compared (Fig. 3).

**Hexane exposure and clogging of Å-slit channels of varied heights**

To start with, we examine a five-layer graphite device ($h$ ~ 1.7 nm) deliberately exposed to hexane. Before introducing the hexane, the channels show high helium flow ~ $5.6 \times 10^{-15}$ mol s$^{-1}$ mbar$^{-1}$ μm normalized per micron channel length, and per mbar pressure difference (Fig. 3b). Upon introduction of hexane into the channels, the helium flow dropped more than three orders of magnitude going below the detection limit of mass spectrometer. Let us note that the 5-layer channel (~ 1.7 nm) is large enough for hexane (kinetic diameter, ~ 0.42 nm) to enter. After the hexane is evacuated in the chamber, the following He measurement through the channels (first He flush) shows a distinct increase of the gas conductance occurring above ~100 mbar pressure (Fig. 3a). The revival of the channels continues with the second helium flush and stays at a stable gas conductance of ~ $4 \times 10^{-15}$ mol s$^{-1}$ mbar$^{-1}$ μm. A reduction of ~ 25 % in the He flow relative to the pre-hexane exposure hints a small portion of hexane remaining inside the channels. With further helium flushes, the adsorbed hexane could not be removed, indicating a limited unclogging efficiency of the helium flush method. Further desorption of hexane was done by a heat treatment of 150 °C for 20 minutes (purple curve in Fig. 3a), recovering the gas flow. Additional measurements of hexane exposure on graphite and hBN devices with same channel heights are shown in supporting information Fig. S7 and Fig. S8. Let us note that the striking behaviour of the recovery, the steps of increased gas flow in a dynamic fashion after first He flush (blue scatter in Fig. 3a), is not captured in other devices with the same channel height (in Fig. S7 and Fig. S8). Probably, the revival process is transient and may depend on several factors that it cannot be experimentally recorded every time.

Next, we investigate the hexane contamination when the hexane molecule size is comparable to the channel size, in this case, bilayer channel device (h ~ 0.7 nm). To our surprise, complete blockage is caused by hexane and we cannot revive the channels neither by helium flushes nor by heating (Fig. 3c and 3d). Further annealing at high temperatures also did not achieve the revival. We performed the same experiment on monolayer channels ($h$ ~ 0.4 nm) which are thinner than the size of the hexane molecule. An initial He conductance of monolayer channel ~ $2.2 \times 10^{-17}$ mol s$^{-1}$ mbar$^{-1}$ µm, reduced to about ~ $2 \times 10^{-17}$ mol s$^{-1}$ mbar$^{-1}$ µm upon exposure to hexane, and it can be totally revived by heating (Fig. 3e and 3f).

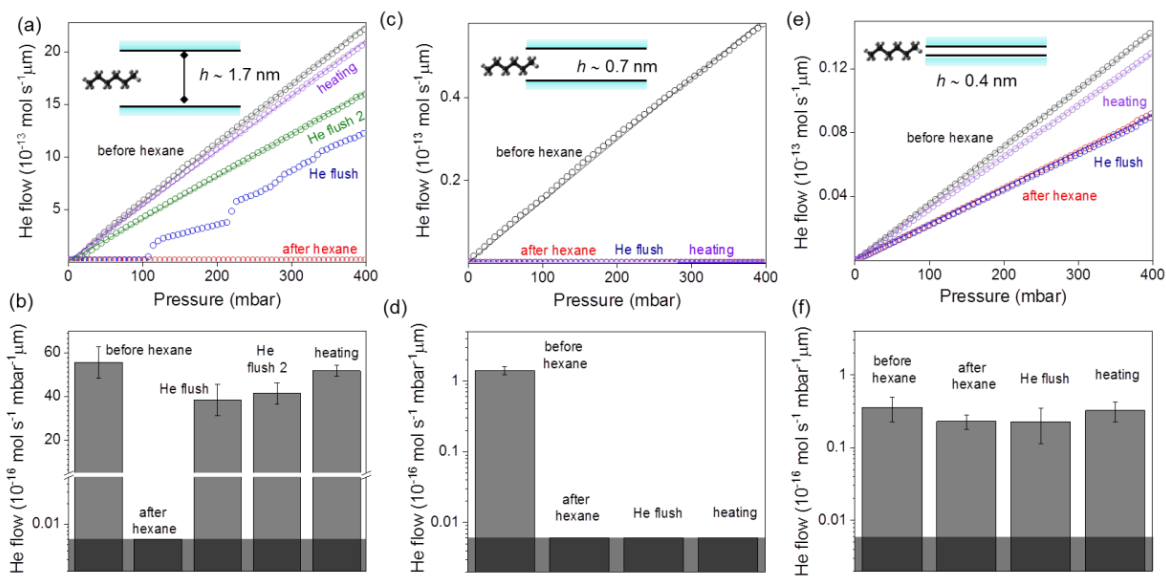

**Fig. 3** Comparison of helium leak rate before and after exposure to hexane through various graphite Å-channel devices with heights, (a) $h$ ~ 1.7 nm, (b) $h$ ~ 0.7 nm, and (c) $h$ ~ 0.4 nm. The insets show the schematics depicting relative size of hexane molecule to the channel in each case. The bar graphs in (b), (d) and (f) are obtained from (a), (c) and (e) and represent normalized He flow per unit pressure. Grey shaded area indicates the limit of detection. All the graphs represent the flows normalized per single channel, and per µm length of the channel. Error bars are from two measurements on the same device, and where there was only measurement (e.g., He flush) it represents uncertainty in the best fit to the measured data.

**The effect of channel height on the interaction of hexane in the Å-slit channels**

Let us discuss the plausible reasons behind the observed phenomenon. Given the size of hexane, it is reasonable to expect it to be rejected from entering the monolayer channels due to the size restriction, as observed in experiments. It has to be noted that monolayer channels possess clean surfaces due to proximity of the top and bottom layers expelling the contaminants outside from self-cleansing effect[37], and can hinder the hydrocarbons from entering. Small reduction in the flow observed for the monolayer channels after exposure to hexane could be possibly due to the trace layer of hexane on the entry blocking the He, and this was easily revived by flushing and heat treatment. With respect to the bilayer channels, they have a relatively close size match between the hexane molecule kinetic diameter and channel height, which could lead to a tight confinement imposing a strong interaction of the channel walls with the long alkyl chain. We observed similar trends using channels with hBN walls (gas flow measurements shown in Fig S8). Let us recall that this principle of close size match between molecules and the pores is explored for preferred adsorption of water and hydrocarbons, combined with capillary condensation at lower than saturation pressures into molecular sieves[42], hydrogen capture into porous systems[43], activated carbon for removal of volatile organic compounds[44].

The ease of clogging and unclogging of channels with varying heights differs and can be largely linked to the strength of the hydrocarbon molecule-surface interaction. Further theoretical studies are required to quantitatively analyze the interaction strength. Here our experimental methods for removing hexane through helium flush and heat treatment can qualitatively compare the hydrocarbon-nanochannel interactions. Helium flush could slightly remove the adsorbed hexane molecules but with limited effect. On the other hand, heating the freshly clogged devices to 150 °C, we could revive the channels by desorbing the adsorbed hexane. Those channels ($h > 2$ nm) which can hardly be recovered by heating to 150 °C, may contain a strong interaction developed between the hydrocarbons and channel surface. Additionally, the severity of clogging increased with the time, as can be seen from the airborne contamination discussed in Fig. 1 and Fig. 2. Thermal annealing at high temperatures has been reported by several groups to decontaminate/desorb hydrocarbons on surfaces of 2D materials, which we also observe here for Å-channels. When the channel size is much larger (e.g., in the case of $h \sim 11$ nm), it proved difficult to revive the pores which may indicate possible formation of hydrocarbon and polymeric clusters which can interact with the channel surface rather than the molecular-level interaction of hydrocarbons.

## Conclusions

In conclusion, we report the influence of the hydrocarbons in a confined channel and discusses the clogging and revival by intentionally exposing hydrocarbon contaminants into the nanochannels. Our experiments highlight that the confining dimension (i.e., channel height in the slits) has a dispensable effect on the severity of the hydrocarbon contamination. Monolayer channel shows a robust stability which can be attributed to their clean surfaces. In contrast, a vigorous confinement from the channel with size closely matching that of the molecular size of the hydrocarbons results in a strengthened surface adsorption, making the recovery and decontamination highly difficult. The proposed method - helium flow measurement - to quantitatively examine the ease of channel clogging and unclogging can benefit several applications of membranes such as water desalination and gas separation in assessing the pore cleanliness and stability.

**Experimental section**

**Fabrication of Å-slit channels**

We fabricated the Å-slit channels by vertically stacking 2D crystals over each other in a layer-by-layer fashion as reported previously[38] (detailed fabrication process flow is given in Supporting information Fig. S1). The device is composed of three layers. The bottom and top 2D crystals in the fabrication of channel devices are chosen to be either graphite or hBN whereas the spacer 2D crystals are always monolayer or few layer thick graphene. The fabrication began with making a free-standing $SiN_x$ membrane (~100 μm × 100 μm) supported over Si substrate. With an additional photolithography step and a reaction ion etching, a rectangular hole of 3 μm × 25 μm was then made on the $SiN_x$ membrane. Bottom, spacer and top layer are transferred one over the other to cover the hole. Spacer layer is typically made up of ~130 (±10) nm wide graphene strips with a ~130 (±10) nm separation using an electron-beam lithography and oxygen plasma etching. A gold patch was made over the tricrystal stack to prevent the sagging of top crystal thin edges into the channels and to use it as a mask to define the channel length. Thus, a 2D channel is made in the tri-crystal stack where the channel length ($L$) is defined by the edge of the metal mask to the edge of the rectangular hole, the channel width ($w$) in our channels is always 130 nm, and the channel height ($h$) is defined by the spacer 2D crystal thickness (mono to few layer thick graphene). The channel width and height are acquired from atomic force microscopy (AFM) (Fig. S9). The AFM images of spacer stripes show their cleanliness, and it is a check point to select the suitable spacers which proceed further for channel fabrication (Fig. S9 and Table S1). After transfer of each layer, the Å-channel devices were annealed in a furnace at 400 °C for 4 hours under a flow of $H_2$/Ar gas mixture. The devices were stores in activated charcoal, please see supporting information S3 for further details about storage.

**Cross-sectional imaging of Å-slit channels**

For the cross-sectional imaging, thin cross-sectional lamellae were made by using in situ lift-out procedure. Perpendicular to the Å-slits' axis, lamellae were cut by high-precision site-specific milling in Helios Nanolab DualBeam 660, incorporating both scanning electron microscope and focused ion-beam columns. To weld the lamella to a micromanipulator, platinum was deposited platinum using the ion beam, enabling its lift-off from the substrate. After transferring to a specialist OmniProbe TEM grid, the lamellae were further thinned to less than 100 nm and polished to electron transparency, using 5-kV and subsequently 2-kV ion milling. High-resolution STEM and HAADF images were acquired in an aberration-corrected microscope (FEI Titan G2 80-200 kV) using a probe convergence angle of 21 mrad, a HAADF inner angle of 48 mrad and a probe current of about 70 pA. The electron beam was aligned parallel to the Å-slits, using the relevant Kikuchi bands of the silicon substrate and the assembled 2D crystals.

**Helium flow measurement**

A leak test was employed to measure the gas flow through the nanochannels of the devices and to verify the clogging and unclogging of the channels. We use a helium leak detector, and the schematic of the setup is depicted in supporting information Fig. S2. It is a two-chamber assembly where one chamber is connected to a voltage control valve to alter helium gas flow, a pressure gauge, a scroll pump to evacuate the system, and a hydrocarbon injecting valve with a swagelok that allows introduction of hexane into the system. The Å-slit channel device is clamped in a customised sample holder which is then held between the two chambers and secured with vacuum O-rings, such that the channels are the only pathway for mass transport. Both ends of the sample holder are connected to the scroll pump that helps to maintain the chambers in vacuum down to $10^{-2}$ mbar. Through a valve, the helium gas was let into the top chamber and permeate through the Å-slit channels to reach the bottom chamber which is connected to a mass spectrometer (INFICON UL200) to measure the flow rate of helium gas. All the measurements are conducted at room temperature of 298 K. For further details about the measurement setup, please refer to supporting information S2.

To evaluate the helium measurement setup, we conducted several trials on control substrates which are blank $SiN_x$/Si wafers, and the reference devices with $SiN_x$ membranes have tri-crystal stack but without any channels in the spacer layer. There is no notable helium flow in both control and reference devices proving that the system is well-sealed.

**Channel reviving process**

In this study, we employed three methods to unclog the channels. One is helium flushing method in which the chamber above the entrance of the channel is filled with helium gas of ~ 1 bar whereas the exit side of the channel faces vacuum. By continuous flushing for 10 minutes, the helium gas could revive the channels. Second one is a heating process where the device is placed on a hot plate heating at 150 °C for 20 minutes to desorb the hexane. While the above two were used in the case of deliberate exposure of channels to hexane, the third method of high temperature annealing in 10% $H_2$/Argon at 400 °C for 4 hours, is employed for routine cleaning of channels from both airborne contamination as well as during fabrication to remove polymer residues.

# Conflicts of interest

The authors declare that there are no conflicting interests.

## Acknowledgements

**Funding:** B.R. acknowledges the funding from the European Union's H2020 Framework Programme/ERC Starting Grant Agreement no. 852674 – AngstroCAP, Royal Society University Research Fellowship URF\R1\180127 and enhancement award RGF\EA\181000, EPSRC Grant EP/R013063/1. A.K. acknowledges Ramsay Memorial Fellowship, the Royal Society research grant RGS\R2\202036.

## Author contributions

B.R. designed and directed the project. A.K., Y.Y., fabricated and characterized the devices. R.Q. characterized the spacers. A.R. and S.H provided STEM measurements. R.S. and S.G. conducted gas measurements. R.S., S.G., A.B. analyzed airborne contamination data. R.S., A.K., B.R. performed hexane contamination experiments and analysis. Y.Y., B.R., A.K., wrote the manuscript with inputs from R.S. All authors contributed to discussions.

Supporting information

# Hydrocarbon Contamination in Angström-scale Channels

Ravalika Sajja,[a,b,†] Yi You,[a,b,†] Rongrong Qi,[a,b] Goutham Solleti,[a,b] Ankit Bhardwaj,[a,b] Alexander Rakowski,[c] Sarah Haigh,[c] Ashok Keerthi,[b,d] Boya Radha,[a,b,]*

[a]Department of Physics and Astronomy, School of Natural Sciences, The University of Manchester, Oxford Road, Manchester M13 9PL, United Kingdom

[b]National Graphene Institute, The University of Manchester, Oxford Road, Manchester M13 9PL, United Kingdom

[c]Department of Materials, School of Natural Sciences, The University of Manchester, Oxford Road, Manchester M13 9PL, United Kingdom

[d]Department of Chemistry, School of Natural Sciences, The University of Manchester, Oxford Road, Manchester M13 9PL, United Kingdom

[†] These authors contributed equally.

*correspondence to: radha.boya@manchester.ac.uk

**Contents:**

**S1. Fabrication of Å-slit channel devices**

**S2. Helium flow measurements**

**S3. Storage and revival of Å-slit channels**

**S4. Airborne contamination in monolayer graphite device**

**S5. Hexane exposure and clogging of graphite Å-slit channels**

**S6. Hexane exposure on a control aperture**

**S7. Hexane exposure and clogging of additional bilayer and five-layer graphite devices**

**S8. Hexane exposure and clogging of hBN Å-slit channels**

**S9. Atomic force microscopy of spacers with varied heights**

**S10. References**

## S1. Fabrication of Å-slit channel devices

Graphite bulk crystals are obtained from Manchester Nanomaterials. Two dimensional (2D) crystals were mechanically exfoliated using scotch tape to expose a fresh crystal on SiO$_2$/Si wafers with ~290 nm thickness of SiO$_2$. Photoresist (S1813) and developer (MF319) for photolithography and polymethyl methacrylate (950k) resist for electron beam lithography (EBL) were purchased from Microposit®. Reactive ion etching (RIE) was used for dry etching silicon nitride (SiN$_x$) and 2D crystals. Micron size rectangular holes in the SiN$_x$ membrane were etched using RIE with a mixture of SF$_6$ and CHF$_3$ gases. Graphite was etched using RIE with oxygen gas.

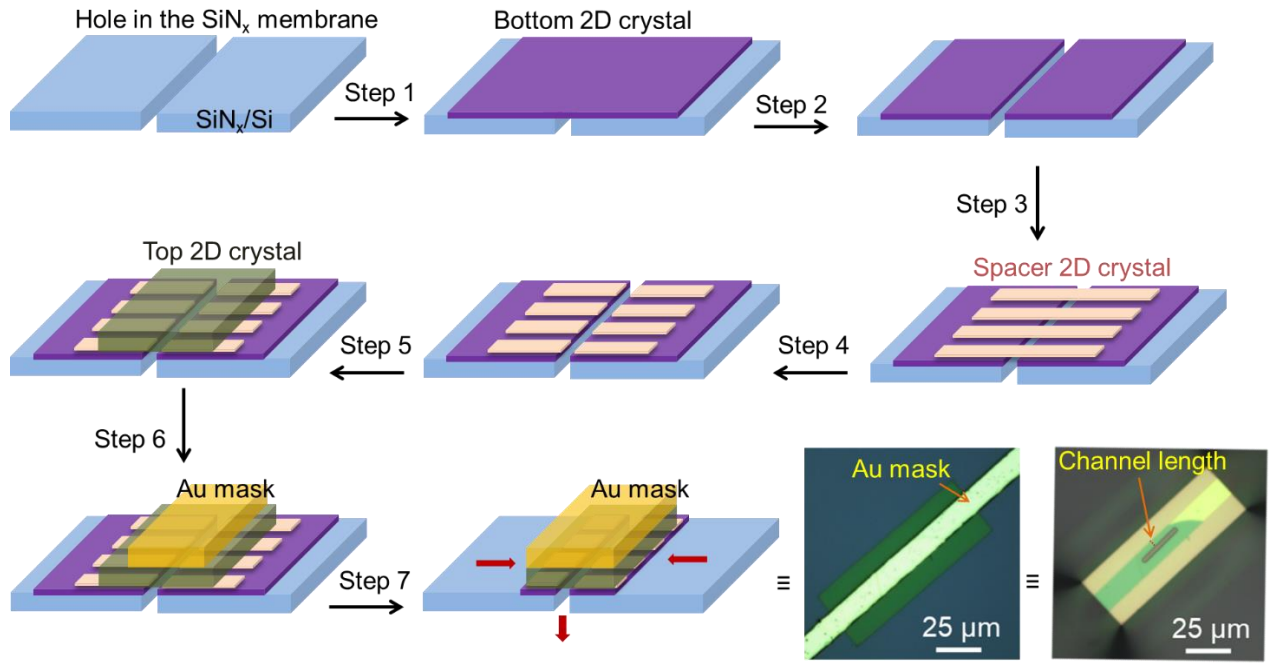

**Figure S1.** Device fabrication flow-chart for making Å-channels. Devices are made using previously reported nanofabrication procedures and the fabrication steps are illustrated with the black arrows.[1] The bottom and top 2D crystals in the fabrication of channel devices are chosen to be either graphite or hexagonal boron nitride (hBN) whereas the spacer 2D crystals are always monolayer or few layer think graphene. **Step 1:** A long rectangular hole in a freestanding silicon nitride (SiN$_x$) membrane (~ 500 nm thick) was covered with the bottom 2D crystal. **Step 2:** The rectangular hole was projected on the bottom crystal layer using RIE (oxygen plasma for graphite, CHF$_3$/oxygen plasma for hBN) from the backside of SiN$_x$ membrane. **Step 3:** The spacer 2D crystal, pre-patterned by EBL using polymethyl methacrylate (PMMA) as a resist and exposed to oxygen plasma to make parallel long stripes with ~ 130 (±10) nm wide and ~ 130 (±10) nm spacing was transferred on to this projected the aperture in the bottom crystal. **Step 4:** This spacer of graphite strips was etched from backside of the membrane to remove exposed graphite on the hole area. **Step 5:** The hole was covered with a relatively thick (~ 200 nm) top 2D crystal. **Step 6:** In some of the devices where top crystals had thin or uneven edges, a metal mask was deposited after a photolithography patterning. **Step 7**: RIE was employed to remove the unmasked thin edges of the top crystal to open channels' entries. After each 2D crystal (bottom, spacer and top) transfer,

the SiN$_x$/Si wafer chip with rectangular hole was annealed in 10% hydrogen-in-argon at 400 °C for 4 hours. The final devices contained between ~ 100 and ~ 2000 channels in total on either side of the rectangular hole.

## S2. Helium flow measurements

The device with Å-channels on SiN$_x$/Si wafer was clamped between O-rings to separate two oil-free chambers (loading and vacuum chambers) as depicted in Fig. S2. In this setup, the only pathway between the chambers is through the channels. Both the chambers were evacuated using a bypass loop connected to a vacuum pump. The chambers were evacuated before every experiment. The vacuum (bottom) chamber was maintained at a pressure of around $10^{-6}$ bar and connected to a mass spectrometer. The loading (top) chamber was equipped with an electrically controlled dosing valve that provided the controlled pressure $P$ inside, which was monitored by a pressure gauge. Helium gas was released into the loading chambers and the applied pressure was varied in a controlled fashion (~ 6 mbar/s) using an electrically controlled gas dosing valve (VAT Group). Our Å-slit channel devices were sufficiently robust to withstand applied pressure up to two bar.

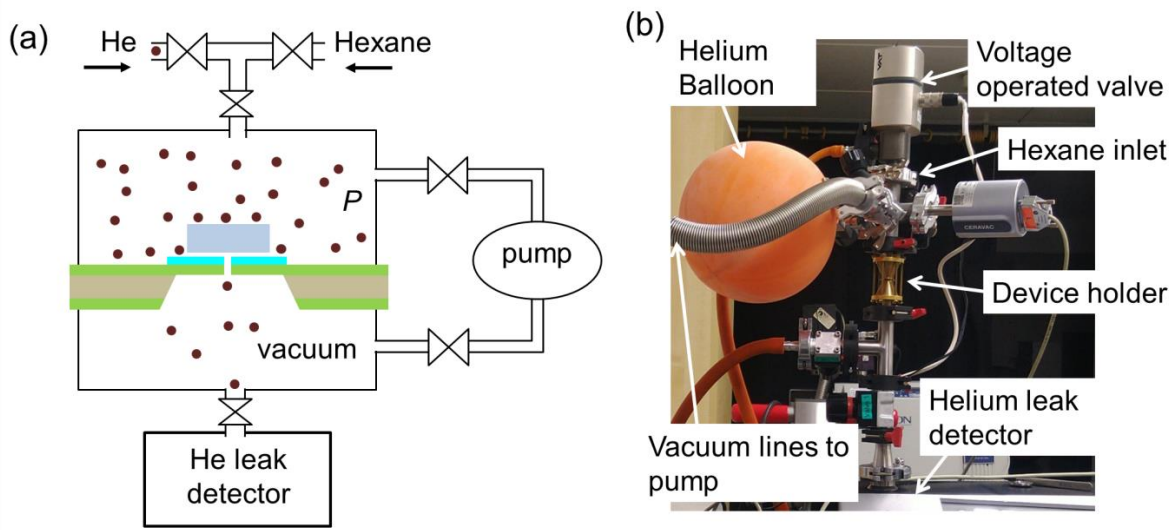

**Figure S2.** (a) Schematic representation of helium gas flow measurements using Å-channel devices. (b) The picture of our experimental setup with helium leak detector and voltage operated valve to selectively inject helium gas and hexane in to the loading chamber.

## S3. Storage and revival of Å-slit channels

After the devices were made using the nanofabrication steps (described above in Fig. S1) in the cleanroom (class 100), they were stored at ambient conditions on the laboratory bench top. These channels (in fact any 2D materials surfaces) can be contaminated with air-borne hydrocarbons in the atmosphere. To protect from the hydrocarbon adsorption and resulting clogging of channels, the devices were stored in activated carbon (Merck, "charcoal activated for analysis"). For measurements, the devices were taken out, washed with water, and IPA and dried under flow of nitrogen gas. Most of the charcoal particles were removed in this cleaning procedure but few particles were left on the devices as depicted in Fig. S3b. We have found that storage in water (in combination with 400 °C annealing) also enables the channels to remain open for few weeks to months.

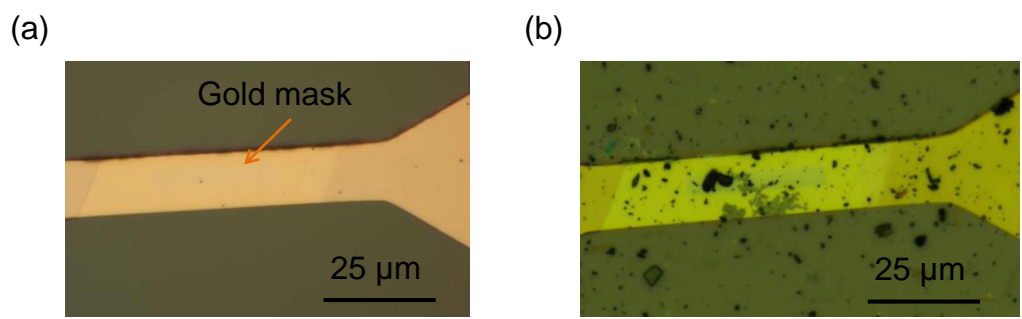

**Figure S3.** Optical image of as fabricated Å-channel device before (a) and image of the same device stored in charcoal for a year (b).

## S4. Airborne contamination in monolayer graphite device

We monitored the airborne contamination of a freshly fabricated monolayer graphite device (channel height h ~ 0.4 nm). The channel shows some variation of the helium permeance monitored for 270 days and could be revived by combination of storage in charcoal and high temperature (400 °C) annealing.

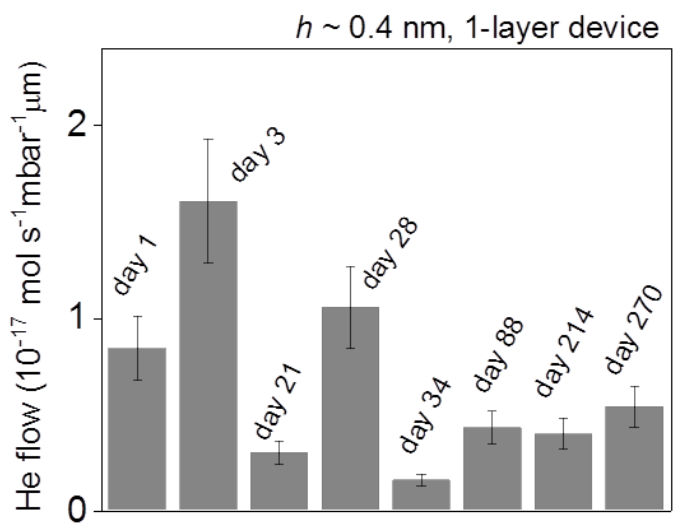

**Figure S4**. Helium flow through a Å-channels made with monolayer graphene spacer (height, $h$ ~ 0.4 nm), top and bottom graphite walls. The bars represent He flows arbitrarily checked on different days through the same device, when it was stored under charcoal. The device was annealed before each measurement.

## S5. Hexane exposure and clogging of graphite Å-slit channels

The Å-channel graphite device (both bottom and top layers made by graphite crystals) were exposed to hexane in the same experimental set-up used for He gas flows as described in the above section (S2) with slight modification. Additional valve and liquid hexane reservoir with < 1 mL capacity were attached to loading chamber (Fig. S2). After the initial He flow measurements through Å-channels, both chambers were evacuated, and hexane was released in to the input chamber. The channel was exposed to vaporized hexane for 60 seconds while the applied pressure in the top input chamber raised up to ~ 200 mbar at temperature T ~ 26 (±1) °C, which is approximately equal to the vapor pressure of hexane and this pressure is sensitive to the experimental temperature. After the chamber was evacuated to remove the hexane for 10 minutes, helium conductance data was measured. Helium flow was recorded using Helium leak detector while input of Helium from a balloon was controlled by electrical voltage gated valve. Opening the electrical gate valve for introducing Helium into input chamber, and the leak detector measurements were started simultaneously. LabVIEW program was used to interface both the pressure and leak rate measurements which are recorded every second. Blank silicon substrates were used to demonstrate the increase in the vapour pressure of hexane with increase in temperature.

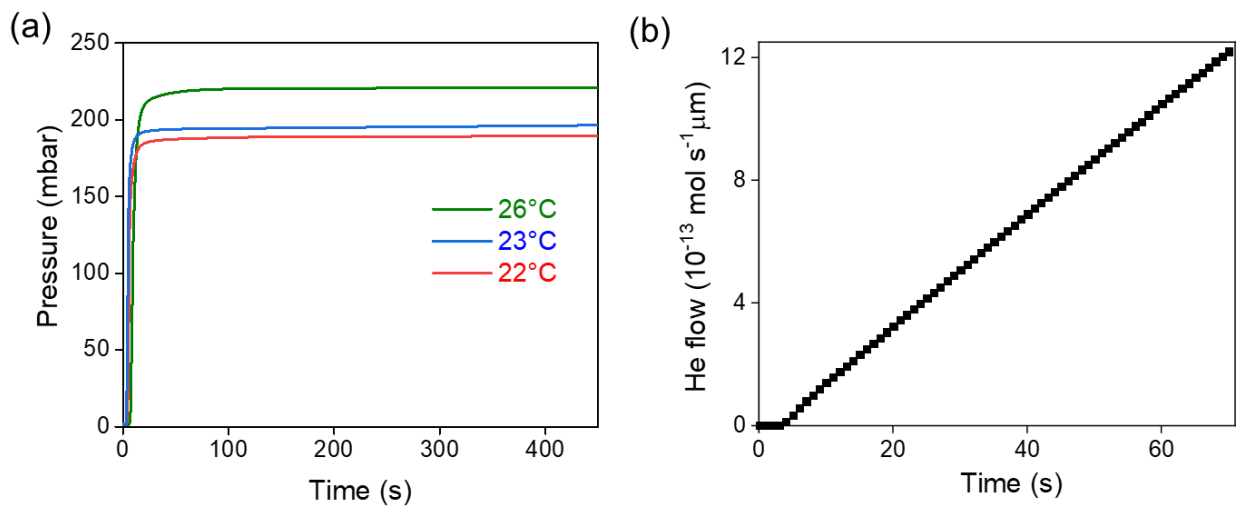

**Figure S5.** (a) Effect of temperature on the vapour pressure of hexane. (b) Time trace plot of the helium flow for 5-layer graphite device.

## S6. Hexane exposure on a control aperture

Control devices, SiN$_x$ membranes without channels, were used to check the measurement set-up thoroughly for possible leaks. These control devices exhibited no noticeable He leak which confirms that the Å-channels were the only possible permeation path. The experimental accuracy of our helium flow measurements was tested using the reference devices containing few nanometres to micrometre sized circular apertures made in free-standing graphene on SiN$_x$ membranes. We did the same sequence of steps, exposure to hexane, recovery by He flush and heat treatment of the substrate. We observe no discernible change in the flow through the aperture, confirming that the measurement setup does not show any flow reduction due to hydrocarbon adsorption on the chamber components or O-rings (Fig. S6).

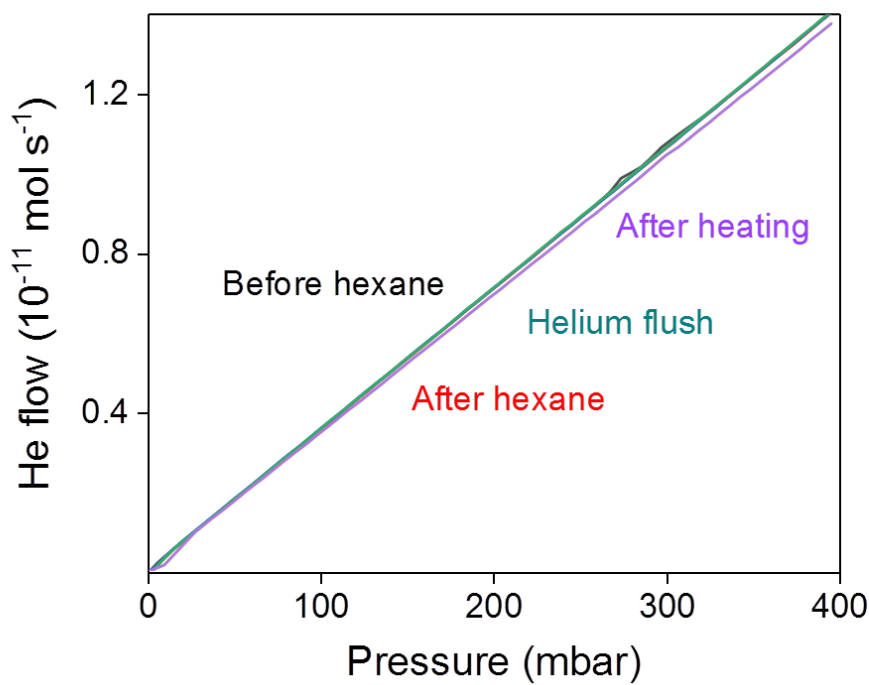

**Figure S6.** Hexane exposure on an aperture of diameter, ∼ 60 nm made in a free-standing graphene layer. The Helium flow remains constant after exposure to hexane and further recovery steps of He flush and heat treatment. This validates our experimental method that the observed changes of flow in the case of Å-channels are mainly due to the channel clogging and recovery.

## S7. Hexane exposure and clogging of additional bilayer and five-layer graphite devices

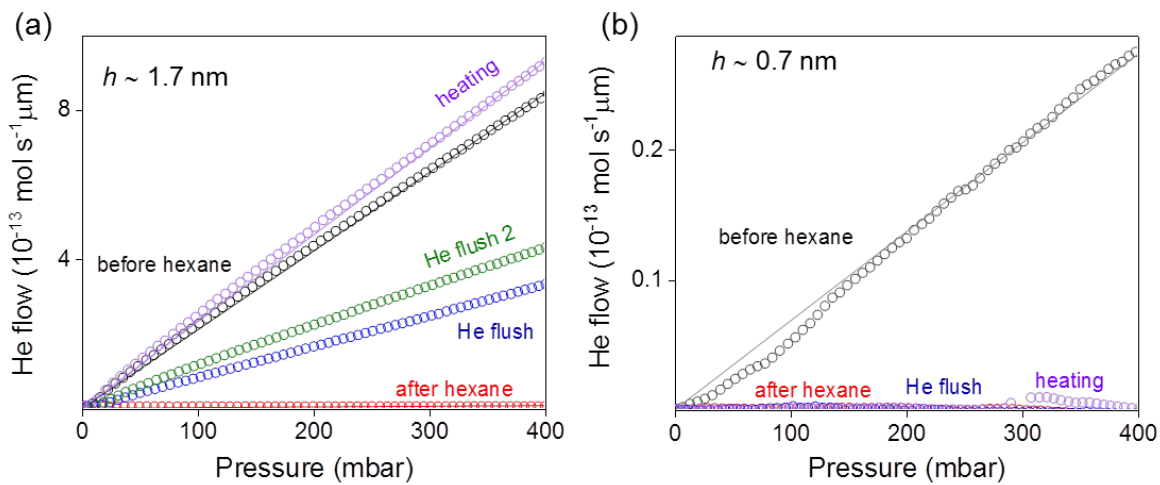

**Figure S7.** Comparison of helium leak rate before and after exposure to hexane through various graphite Å-channel devices with heights, (a) $h \sim 1.7$ nm, (b) $h \sim 0.7$ nm. All the graphs represent the flows normalized per single channel, and per µm length of the channel.

## S8. Hexane exposure and clogging of hBN Å-slit channels

The Å-channel hBN channels (both bottom and top layers made by hBN crystals) were exposed to hexane in the same experimental set-up used for He gas flows in graphite channel device (S7).

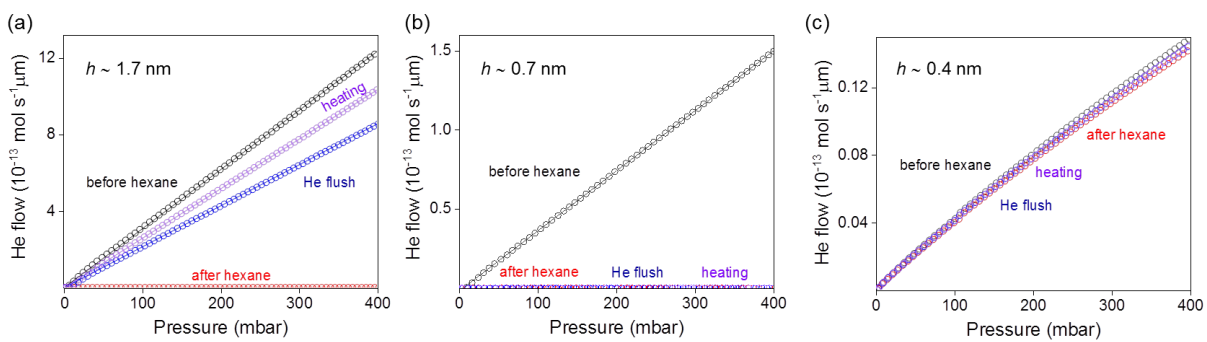

**Figure S8.** Comparison of helium leak rate before and after exposure to hexane through various hBN Å-channel devices with heights, (a) $h \sim 1.7$ nm, (b) $h \sim 0.7$ nm, and (c) $h \sim 0.4$ nm. All the graphs represent the flows normalized per single channel, and per µm length of the channel.

## S9. Atomic force microscopy of spacers with varied heights

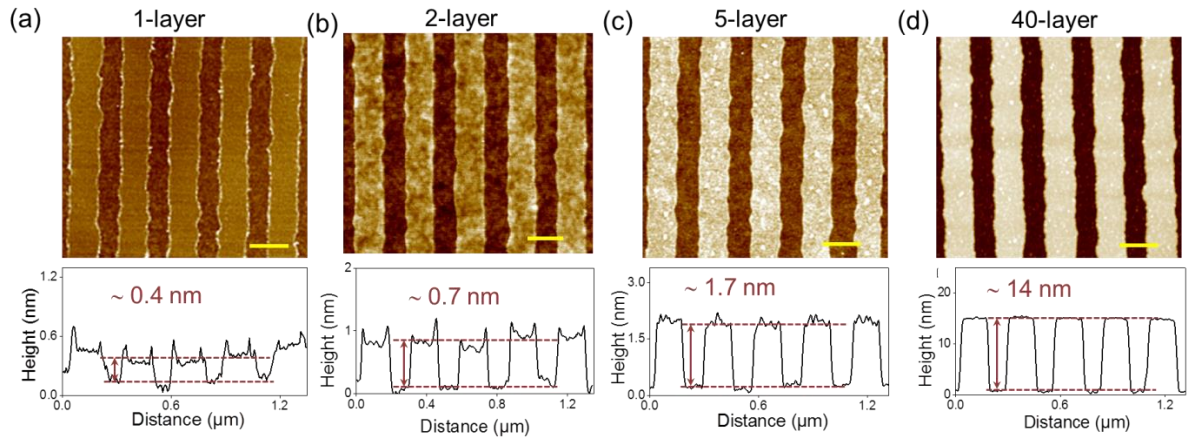

**Figure S9.** Atomic force micrographs and height profile of (a) monolayer spacer, (b) bilayer spacer, (c) five-layer spacer and (d) forty-layer spacer. Scale bar, 200 nm.

**Table S1. Surface roughness of monolayer, bilayer, five-layer and forty-layer graphene spacers.**

| Surface roughness | 1-layer spacer | 2-layer spacer | 5-layer spacer | 40-layer spacer |
|---|---|---|---|---|
| $R_q$ | 0.6 ± 0.1 Å | 1.9 ± 0.2 Å | 2.5 ± 0.4 Å | 4.2 ± 0.3 Å |
| $R_a$ | 0.5 ± 0.1 Å | 1.5 ± 0.1 Å | 1.8 ± 0.2 Å | 3.2 ± 0.3 Å |

We analysed the surface roughness (e.g. root mean square roughness $R_q$ and mean roughness $R_a$) based on the AFM images above (Fig. S9). For each spacer, we recorded the surface roughness value of individual spacer strip and the roughness value was obtained by averaging the data of at least five spacer strips.